\documentclass{article}
\usepackage[left=2.2cm, right=2.2cm, top=2.2cm]{geometry}
\usepackage{amssymb,amsmath,verbatim,mathtools,needspace,enumitem,etoolbox,graphicx,physics,microtype,afterpage,xspace,tabularx,lmodern,multirow}
\usepackage{authblk}
\usepackage{gensymb}
\usepackage[utf8]{inputenc}
\usepackage[usenames,dvipsnames]{xcolor}
\definecolor{linkcolor}{rgb}{0.0,0.3,0.5}
\definecolor{romared}{RGB}{142,0,28}
\usepackage[unicode, colorlinks=true, linkcolor=linkcolor, citecolor=blue, filecolor=linkcolor, urlcolor=linkcolor, linktocpage, breaklinks]{hyperref}
\usepackage[all]{hypcap}
\usepackage[T1]{fontenc}
\usepackage{subcaption}
\usepackage{url}

\newcommand{\mituni}{Center for Theoretical Physics, Massachusetts Institute of Technology, 77 Massachusetts Ave, Cambridge, MA 02139, USA}

\newcommand{\qmul}{Geometry, Analysis and Gravitation, School of Mathematical Sciences, Queen Mary University of London,
Mile End Road, London E1 4NS, United Kingdom}
\newcommand{\kcl}{Theoretical Particle Physics and Cosmology Group, Physics Department, King's College London, Strand, London WC2R 2LS, United Kingdom}
\newcommand{\bilbao}{Department of Physics, University of Basque Country, UPV/EHU, 48080, Bilbao, Spain}
\newcommand{\jhu}{Department of Physics and Astronomy, Johns Hopkins University, Baltimore, MD 21218, USA}
\newcommand{\uiuc}{Department of Physics, University of Illinois Urbana-Champaign, Urbana, IL 61801, USA}
\newcommand{\camb}{Information Services, University of Cambridge, Roger Needham Building, 7 JJ Thomson Avenue, Cambridge, CB3 0WA, United Kingdom}
\newcommand{\sto}{Institute for Advanced Computational Science, Stony Brook University, NY 11794 USA}
\newcommand{\leu}{Institute for Theoretical Physics, KU Leuven, Celestijnenlaan 200D, B-3001 Leuven, Belgium}
\newcommand{\leuv}{Leuven Gravity Institute, KU Leuven, Celestijnenlaan 200D, B-3001 Leuven, Belgium}

\title{\textbf{\texttt{GRTresna}: An open-source code to solve the initial data constraints in numerical relativity}}
\author[1]{Josu C. Aurrekoetxea}
\author[2]{Sam E. Brady}
\author[3,4]{Llibert Arest\'e-Sal\'o}
\author[5]{Jamie Bamber}
\author[6]{Liina Chung-Jukko}
\author[2]{Katy Clough}
\author[6]{Eloy de Jong}
\author[7]{Matthew Elley}
\author[2]{Pau Figueras}
\author[8]{Thomas Helfer}
\author[6]{Eugene A. Lim}
\author[9]{Miren Radia}
\author[2]{Areef Waeming}
\author[10]{Zipeng Wang}

\date{}

\affil[1]{\mituni}
\affil[2]{\qmul}
\affil[3]{\leu}
\affil[4]{\leuv}
\affil[5]{\uiuc}
\affil[6]{\kcl}
\affil[7]{\bilbao}
\affil[8]{\sto}
\affil[9]{\camb}
\affil[10]{\jhu}

\begin{document}

\maketitle

\texttt{GRTresna} is a multigrid solver designed to solve the constraint equations for the initial data required in numerical relativity simulations. In particular, it is focussed on scenarios with fundamental fields around black holes and inhomogeneous cosmological spacetimes. The code is based on the formalism in Aurrekoetxea, Clough \& Lim \cite{Aurrekoetxea:2022mpw} and can be found at \url{https://github.com/GRTLCollaboration/GRTresna}\footnote{We follow the GRTL Collaboration convention of naming codes as GR[Tool], where ``tresna'' means ``tool'' in \textit{Euskara} (Basque) \cite{enwiki:euskara}.}.

\section{Summary}

Numerical relativity (NR) is a tool for the solution of the Einstein Equations, which describe gravity in strong field regimes. The equations can be expressed as a set of coupled partial differential equations (PDEs) for the 10 metric quantities $g_{\mu\nu}$ and their time derivatives $\partial_t g_{\mu\nu}$. NR is primarily focussed on the hyperbolic PDEs that describe their time evolution from an initial data set, but the initial data itself must satisfy a set of four coupled non-linear elliptic PDEs known as the Hamiltonian and momentum constraints. Whilst these constraints can be solved more straightforwardly by making certain assumptions, this significantly restricts the range of physical scenarios that can be studied. A general solver therefore expands the physics that NR evolutions can be used to probe. 

In the ADM form of the Einstein Equations \cite{Arnowitt:1962hi}, we slice the spacetime into 3-dimensional hypersurfaces
\begin{equation}
    ds^2 = -(\alpha^2 - \beta_i\beta^i) dt^2 + 2\beta_i dx^i dt + \gamma_{ij} dx^i dx^j
\end{equation}
where $\alpha$ and $\beta^i$ are the lapse and shift functions, respectively. The Hamiltonian and momentum constraints are expressed as
\begin{align}
\mathcal{H} &\equiv R + K^2-K_{ij}K^{ij}-16\pi \rho = 0\,, \label{eq:Ham} \\
\mathcal{M}_i &\equiv D^j (K_{ij}- \gamma_{ij} K) - 8\pi S_i = 0\,. \label{eq:Mom}
\end{align}
Here, $\gamma_{ij}$ is the 3-metric of the hypersurface, $R$ is the Ricci scalar associated to this metric, and $K_{ij}\sim \partial_t \gamma_{ij}$ is the extrinsic curvature tensor, with $K=\gamma^{ij}K_{ij}$ its trace. The decomposed components of the stress-energy tensor of matter (measured by normal observers) are defined as $\rho = n_\mu\,n_\nu\,T^{\mu\nu}$ and $S_i = -\gamma_{i\mu}\,n_\nu\,T^{\mu\nu}$, where $n_\mu = (-\alpha,0,0,0)$. These equations constitute the set of four PDEs to be solved. There are 16 unknowns: 6 in $\gamma_{ij}$, 6 in $K_{ij}$, 1 in $\rho$ and 3 in $S_i$. Usually the matter configuration is set by the physical scenario, which determines $\rho$ and $S_i$. The constraints only determine 4 quantities, and the remaining 8 (4 of which are physical degrees of freedom, and 4 gauge choices) must be chosen according to physical principles or knowledge about the system. In this short paper, we introduce \texttt{GRTresna}, an open-source code to solve these equations.

The two main methods for finding initial conditions in numerical relativity are the \emph{conformal transverse-traceless} (CTT) and the \emph{conformal thin sandwich} (CTS) approaches.  We refer the reader to the standard NR texts \cite{Alcubierre:2008co,Gourgoulhon:2007ue,Baumgarte:2010ndz,Baumgarte:2021skc,Shibata_book} for more details about these. \texttt{GRTresna} implements two variations of the CTT method recently introduced in Aurrekoetxea, Clough \& Lim \cite{Aurrekoetxea:2022mpw}: the CTTK and CTTK-Hybrid methods, which are particularly well-suited to cases with fundamental fields in the matter content. Documentation about using and modifying \texttt{GRTresna} can be found in the code wiki \url{https://github.com/GRTLCollaboration/GRTresna/wiki}.

\begin{figure}[t]
\includegraphics[width=\textwidth]{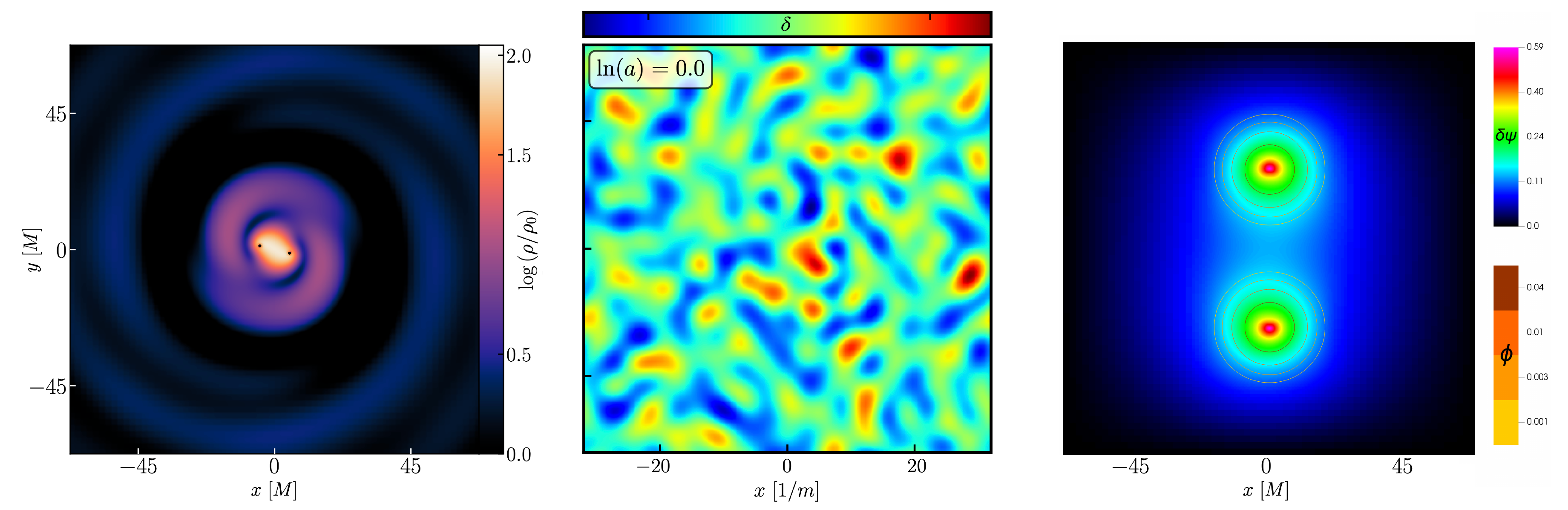}
\caption{Some highlights of work using \texttt{GRTresna} to date: (Left:) Dark matter around binary black holes, from \cite{Bamber:2022pbs,Aurrekoetxea:2023jwk,Aurrekoetxea:2024cqd} (Middle:) Evolution of inflationary perturbations during preheating, from \cite{Aurrekoetxea:2023jwd}. (Right:) Scalar fields around black holes in $4\partial ST$ gravity, from \cite{Brady:2023dgu}. \label{fig:projects}}
\end{figure}

\section{Key features of \texttt{GRTresna}}

The key features of \texttt{GRTresna} are as follows
\begin{itemize}
    \item Flexibility: \texttt{GRTresna} is designed to be extended to various physical scenarios, including different matter types and gravitational theories beyond GR. It currently supports cosmological-type periodic spacetimes and a superposition of two boosted and/or spinning black holes (Bowen-York initial data), with fully general scalar field matter source configurations and the flexibility to adapt to other setups. While scalar fields are the only matter sources included in the current version of the code, the templated methods allow users to easily replace them with other matter types by copying the scalar field implementation and modifying the methods to compute the corresponding energy and momentum densities.
    
    \item Methods: \texttt{GRTresna} incorporates the CTTK and CTTK-Hybrid methods to solve the Hamiltonian and momentum constraints. These methods offer several advantages when dealing with fundamental fields, as discussed in \cite{Aurrekoetxea:2022mpw}. The method code is also templated, so users can easily implement their preferred methods.

    \item Initial conditions: The code supports analytical initial data for the matter fields, as well as the option to read grids and data from an existing HDF5 file. This functionality is especially useful when combined with our code that evolves matter on fixed metric backgrounds \texttt{GRDzhadzha} \cite{Aurrekoetxea:2023fhl}, meaning that we can upgrade the resulting matter configurations to full NR simulations with backreaction.
    
    \item Boundary conditions: The code implements extrapolating, reflective, and periodic boundary conditions, compatible with those in the NR evolution code \texttt{GRChombo} \cite{Andrade:2021rbd,Clough:2015sqa}.

    \item Diagnostics: The code computes the Hamiltonian and momentum constraint errors at each iteration step, and outputs the norm of these values across the grid to a text file.

    \item Compatibility: As \texttt{GRTresna} is developed on top of \texttt{Chombo}, the solver is primarily designed to be compatible with \texttt{GRChombo} \cite{Andrade:2021rbd,Clough:2015sqa} and the family of codes developed by the GRTL Collaboration. We provide two examples that integrate directly with existing examples in the GRChombo evolution code (via the output of a checkpoint file for restart at $t=0$), and provide guidance and tools to validate the results. However, the code outputs data in the standard HDF5 data format, which should be straightforward to adapt to other NR codes that support HDF5 input or can be accessed using Python.
\end{itemize}

Other features that are inherited from \texttt{Chombo} include

\begin{itemize}
    
    \item C++ class structure: \texttt{GRTresna} is written in the C++ language, and makes heavy use of object-oriented programming (OOP) and templating.
    
    \item Parallelism: \texttt{GRTresna} uses hybrid OpenMP/MPI parallelism.
    
    \item Adaptive Mesh Refinement: The code inherits the flexible AMR grid structure of Chombo,
    with block-structured Berger-Rigoutsos grid generation \cite{Berger:1991}. The tagging of refinement regions is fully flexible and while it is based on the sources of the elliptic equations by default, other user-defined measures can be defined \cite{Radia:2021smk}.

    \item Fast: The code uses a multigrid method to efficiently reduce errors across a hierarchy of discretizations, enabling the solver to achieve rapid convergence while minimizing computational costs. This makes \texttt{GRTresna} highly optimized for handling the demanding computations of initial data in the presence of AMR.
\end{itemize}

Forthcoming features currently under development include the addition of other methods, in particular the Extended Conformal Thin Sandwich (XCTS) method, non-conformally flat metric data, new matter types including vector fields, the modified scalar-tensor gravity formalism of Brady et. al. \cite{Brady:2023dgu} and dimensional reduction to 2D using the cartoon formalism \cite{Alcubierre:1999ab, Cook:2016soy}.

\section{Statement of need}

There are a number of existing initial data solvers for numerical relativity, most of which are primarily designed to solve for initial conditions in compact object mergers (i.e. neutron stars and black holes). These include \texttt{TwoPunctures} \cite{Ansorg:2004ds}, \texttt{SGRID} \cite{Tichy:2009yr}, \texttt{BAM} \cite{Bruegmann:2006ulg}, \texttt{LORENE} \cite{LORENE,Gourgoulhon:2000nn}, \texttt{Spells} \cite{Pfeiffer:2002wt}, \texttt{SpECTRE} \cite{Vu:2021coj,Vu:2024cgf,deppe_2023_8196313} (\cite{Nee:2024bur} for modified gravity), \texttt{COCAL} \cite{Uryu:2011ky,Tsokaros:2012kp,Tsokaros:2015fea}, \texttt{PCOCAL} \cite{Boukas:2023ckb}, \texttt{Elliptica} \cite{Rashti:2021ihv}, \texttt{NRPyElliptic} \cite{Assumpcao:2021fhq}, \texttt{KADATH}/\texttt{FUKA} \cite{FUKA,Grandclement:2009ju,Papenfort:2021hod}, \texttt{SPHINCS\_ID} \cite{SPHINCSID,Diener:2022hui,Rosswog:2023nnl}, and the solver of East \emph{et al.} \cite{East:2012zn}. Many of these codes, particularly those using spectral methods like \texttt{TwoPunctures} and \texttt{SpECTRE}, provide a higher accuracy in the solution compared to \texttt{GRTresna}, which is limited to second order accuracy by the multigrid method used. They are therefore better suited to initial data for waveform generation where precision is key. \texttt{GRTresna} is, however, designed to be more flexible and general purpose, tackling both cosmological and black hole spacetimes in a range of scenarios beyond GR and the Standard Model.

In particular, to the best of our knowledge, there is no fully general, publicly available initial condition solver for inhomogeneous cosmological spacetimes. One exception is \texttt{FLRWSolver}, developed by Macpherson \emph{et al.} \cite{Macpherson:2016ict} as part of \texttt{the Einstein Toolkit} \cite{Loffler:2011ay}, which specializes in initializing data for cosmological perturbations arising from inflation for studies of late-time cosmology. However, this is limited to only weakly non-linear initial data. \texttt{GRTresna} aims to provide an open-source tool that not only incorporates the general features of existing initial data solvers for compact objects in GR but also extends their capabilities to cosmological spacetimes (see \cite{Aurrekoetxea:2024mdy} for a review of the application of numerical relativity in cosmology). \texttt{GRTresna} is particularly well-suited for fundamental field matter types, such as scalar and vector fields. Its flexible design allows users to implement new solver methods, additional matter types, or extend the code to study theories beyond GR. It is fully compatible with the GRTL Collaboration's ecosystem of codes but can also serve as a complementary tool for generating constraint-satisfying initial data for other numerical relativity codes.

\section{Research projects to date using \texttt{GRTresna}}

The code has already been used successfully to study a range of problems in fundamental physics, including:
\begin{itemize}
    \item The robustness of inflation to inhomogeneities in the scalar field \cite{Aurrekoetxea:2019fhr,Elley:2024alx}.
    \item The formation of oscillons during inflationary preheating \cite{Aurrekoetxea:2023jwd}.
    \item Formation of spinning primordial black holes \cite{deJong:2023gsx}.
    \item The effect of scalar dark matter environments around binary black holes \cite{Bamber:2022pbs,Aurrekoetxea:2023jwk,Aurrekoetxea:2024cqd}.
    \item The general relativistic evolution of polarized Proca stars \cite{Wang:2023tly}.
    \item Solving the initial conditions problem for modified gravity theories \cite{Brady:2023dgu}.
\end{itemize}

\section*{Acknowledgements}

\noindent We thank the GRTL collaboration (\href{www.grtlcollaboration.org}{www.grtlcollaboration.org}) for their support and code development work.
JCA acknowledges funding from the Department of Physics at MIT. SEB is supported by a QMUL Principal studentship. JB acknowledges funding by the National Science Foundation (NSF) Grants PHY-2308242, OAC-2310548 and PHY-2006066 to the
University of Illinois at Urbana-Champaign. KC acknowledges funding from the UKRI Ernest Rutherford Fellowship (grant number ST/V003240/1). KC and PF acknowledge funding from STFC Research Grant ST/X000931/1 (Astronomy at Queen Mary 2023-2026). ME has been supported in part by the PID2021-123703NB-C21 grant funded by MCIN/AEI/10.13039/501100011033/and by ERDF: ``A way of making Europe''; the Basque Government grant (IT-1628-22). EAL acknowledges support from a Leverhulme Trust Research Project Grant. AW acknowledges the support of the Development and Promotion of Science and Technology Talents Project (DPST), the Institute for the Promotion of Teaching Science and Technology (IPST), Thailand. ZW is supported by NSF Grants No. AST-2307146, PHY-2207502, PHY-090003 and PHY-20043, by NASA Grant No. 21-ATP21-0010, by the John Templeton Foundation Grant 62840, by the Simons Foundation, and by the Italian Ministry of Foreign Affairs and International Cooperation grant No. PGR01167.

Development of this code used the DiRAC Memory Intensive services Cosma8 and Cosma7 at Durham University, managed by the Institute for Computational Cosmology on behalf of the STFC DiRAC HPC Facility (www.dirac.ac.uk). The DiRAC service at Durham was funded by BEIS, UKRI and STFC capital funding, Durham University and STFC operations grants. DiRAC is part of the UKRI Digital Research Infrastructure.

\bibliographystyle{vancouver}
\bibliography{main}

\end{document}